\documentclass[11pt]{article}
\usepackage{amssymb,amsmath,amsthm,amscd,latexsym}
\usepackage{mathrsfs}
\usepackage{mathrsfs}
\usepackage{amsfonts}
\usepackage{amsmath}
\usepackage{amssymb}
\usepackage{amscd}
\usepackage[all]{xy}
\usepackage{xypic}
\usepackage{graphicx}
\usepackage{epstopdf}
\usepackage{color}

\renewcommand{\paragraph}{\roman{paragraph}}
\input cyracc.def

\newcommand{\F}{\mathbb{F}}

\begin{document}
%\begin{CJK*}{GBK}{song}\CJKtilde
\title{\bf Linear codes with few weights over $\mathbb{F}_2+u\mathbb{F}_2$
\thanks{This research is supported by National Natural Science Foundation of China (61672036), Technology Foundation for Selected Overseas Chinese Scholar, Ministry of Personnel of China (05015133) and the Open Research Fund of National Mobile Communications Research Laboratory, Southeast University (2015D11) and Key projects of support program for outstanding young talents in Colleges and Universities (gxyqZD2016008).}
}
\author{\small{Minjia Shi, Liqin Qian}\\ \small{School of Mathematical Sciences of Anhui University, China}
\and \small{Patrick Sol\'e}\\ \small{Telecom Paris Tech, France and King Abdulaziz University, Saudi Arabia}
}
\date{}
\maketitle
%\end{CJK*}
%\begin{CJK}{GBK}{song}
{\bf Abstract:} {In this paper, we construct an infinite family of five-weight codes from trace codes over the ring
$R=\mathbb{F}_2+u\mathbb{F}_2$, where $u^2=0.$ The trace codes have the algebraic structure of abelian codes. Their Lee weight is computed by using character sums. Combined with Pless power moments and Newton's Identities, the weight distribution of the Gray image of trace codes was present. Their support structure is determined. An application to secret sharing schemes is given.

%Finally, the specific forms of generating idempotents of quadratic residue codes of length 11 over $F_{3}+vF_{3}+v^{2}F_{3}$ are given to illustrate our results.}


\section{Introduction}

\hspace*{0.6cm}The two-weight codes are an important topic of cryptography and coding theory due to their connections to strongly regular graphs, finite geometries and association schemes \cite{D}. A classical construction of codes over finite fields called trace codes is documented. Many known codes can be produced by this method. However, as was observed in \cite{AE}, the constructions have used cyclic codes.

In a recent interesting paper \cite{SL, SL1, SL4, SL2, SL3}, the notion of trace codes has been extended from fields to rings. We can summarize this reseach program as shown below: \\
\cite{SL}: $L=R_m^*, R=\mathbb{F}_2+u\mathbb{F}_2;$\\
\cite{SL1}: $L=R_m^*, R=\mathbb{F}_2+u\mathbb{F}_2+v\mathbb{F}_2+uv\mathbb{F}_2;$\\
\cite{SL4}: $L=D+u\mathbb{F}_{p^m}, (p-1)|[R_m^*,L], R=\mathbb{F}_p+u\mathbb{F}_p;$\\
\cite{SL2}: $[R_m^*:L]=2, R=\mathbb{F}_p+u\mathbb{F}_p;$\\
\cite{SL3}: $L=R_m^*, R=\mathbb{F}_2+u\mathbb{F}_2+u^2\mathbb{F}_2;$\\
Here, $L$ is called the defining set of trace codes, $R_m$ denotes a $m$-extension of the ring $R$ with $m>1$. In fact, they are part of a general research program where a variety of few weight codes are obtained by varying the base ring and the defining set. In the present paper, we defined a trace code by using a different evaluation map. Compared with \cite{SL}, the definition is new and we obtain a different linear codes. In combination with Pless power moments and Newton's Identities, we investigate the Lee weight distribution of trace codes.\\
\hspace*{0.6cm}The manuscript is organized as follows. Basic notations and definitions are provided in Section 2. Section 3 shows that the codes and their binary images are abelian. In this paper, main result, the Lee weight of these codes, together with some examples, are presented in Section 4. Section 5 determines the minimum distance of the dual codes. Section 6, we obtain the Lee weight distribution of $C_m$ by using Pless power moments. The support structure of binary images and an application to secret sharing schemes are given in Section 7.

\section{Preliminaries}
\hspace*{0.6cm}We consider the local ring $\mathbb{F}_2+u\mathbb{F}_2$ denoted by $R$, with $u^2=0$. For any positive integer $m$, we construct an extension of degree $m$ as $R_m=\mathbb{F}_{2^m}+u\mathbb{F}_{2^m}$ with again $u^2=0.$ It is a local ring with maximal ideals $(u)$. What's more, it is a Frobenius ring. So there is a \emph{Frobenius operator} $F$ which maps $\alpha+\beta u$ onto $\alpha^2+\beta^2u$. The \emph{Trace} function, denoted by $Tr$ is then defined as
$$Tr=\sum_{j=0}^{m-1}F^j.$$
It is immediate to check that $$Tr(\alpha+\beta u)=tr(\alpha)+tr(\beta)u,$$
 for all $\alpha,\beta\in \mathbb{F}_{2^m}$.
Here, the standard trace of $\mathbb{F}_{2^m}$ denoted by $tr()$.

For convenience, let $M$ denotes the maximal ideal of $R_m,$ i.e., $M=(u)=\{\beta u:\beta\in \mathbb{F}_{2^m}\}$.
The group of units in ${\mathcal{R}_m},$ denoted by ${\mathcal{R}}_m^*,$ is $\{\alpha+\beta u:\alpha\in \mathbb{F}_{2^m}^{*},\beta\in \mathbb{F}_{2^m}\}.$ It is easy to check $R_m^*\cong \mathbb{F}_{2^m}^{*}\times \mathbb{F}_{2^m}$ and $|\mathcal{R}_m^\ast|=(2^{m}-1)2^m$. Hence, ${\mathcal{R}}_m^*$ is not a cyclic group and that ${\mathcal{R}_m}={\mathcal{R}}_m^*\cup M$.

 A \textbf{linear code} $C$ over $R$ of length $n$ is an $R$-submodule of $R^n$. If $x=(x_1,x_2,\cdots,x_n)$
 and $y=(y_1,y_2,\cdots,y_n)$ are two elements of  $R^n$, their standard inner product
  is defined by $\langle x,y\rangle=\sum_{i=1}^nx_iy_i$, where the operation is performed in $R$. The {\bf dual code} of $C$ is denoted by $C^\perp$ and
  defined as $C^\perp=\{y\in R^n|\langle x,y\rangle =0, \forall x\in C\}.$

For $x=(x_1,x_2,\ldots,x_n),y=(y_1,y_2,\ldots,y_n)\in \mathbb{F}_2^n,d_H(x,y)=|\{i:x_i\neq y_i\}|$ is called the \textbf{Hamming distance} between $x$ and $y$ and $w_H(x)=d_H(x,0)$, the Hamming weight of $x$. The Hamming weight of a codeword $c=(c_1,c_2,\ldots,c_n)$ of $\mathbb{F}_2^n$ can also be equivalently defined as $w_H(c)=\sum_{i=1}^{n}w_H(c_i)$, where $w_H(c_i)$ equals to 0 if and only if $c_i$ is a zero element.

 For any $a=\alpha+\beta u\in R$, we define the \textbf{Gray map} $\Phi :R \rightarrow \mathbb{F}_2^{2}$, $\Phi(\alpha+\beta u)=(\beta,\alpha+\beta)$, where $\alpha,\beta\in \mathbb{F}_{2}$. This map can be extended to $R^n$ in the natural way \cite{SL}. From the definition of Gray map, we know that $\Phi$ is bijection and linear. Then $\Phi$ is a weight-preserving map from ($R^n$, Lee weight) to ($\mathbb{F}_2^{2n}$, Hamming weight), that is, $w_L(x)=w_H(\Phi(x)), x\in R^n$.

 Given a finite abelian group $G,$ a code over $R$ is said to be {\bf abelian} if it is an ideal of the group ring $R[G].$ In other words, the coordinates of $C$ are indexed by elements of $G$ and $G$ acts regularly on this set. In the special case when $G$ is cyclic, the code is a cyclic code in the usual sense \cite{MJ}.

\section{ Symmetry }
For $a,b\in \mathcal{R}_m$, we define the vector $Ev(a,b)$ by the following evaluation map: $$Ev(a,b)=(Tr(ax+bx^3))_{x\in \mathcal{R}_m^*}.$$ Define the code $C_m$ by the formula $C_m=\{Ev(a,b)|a,b\in \mathcal{R}_m\}$. Thus $C_m$ is a code of length $|\mathcal{R}_m^\ast|$ over $R$. \\
\noindent\textbf{Proposition 3.1} The group of units $\mathcal{R}_m^*$ acts regularly on the coordinates of $C_m.$\\
{\bf Proof.}
For any $v',u' \in \mathcal{R}_m^*$ the change of variables $ x\mapsto (u'/v')x$ permutes the coordinates of $C_m,$ and maps $v'$ to $u'.$ Such a permutation is unique, given $v',u'.$ \qed

The code $C_m$ is thus an {\em abelian code} with respect to the group $\mathcal{R}_m^*.$ In other words, it is an ideal of the group ring $R[\mathcal{R}_m^*].$ As observed in the previous section, $\mathcal{R}_m^*$ is not a cyclic group, and thus $C_m$ may be not cyclic. The next result shows that its binary image is also abelian.\\
\noindent\textbf{Proposition 3.2} A finite group of size $2|\mathcal{R}_m^*|$ acts regularly on the coordinates of $\Phi(C_m).$\\
{\bf Proof.}
It is similar to the proof in \cite{SL}, and we omit it here.\qed\\

\section{ The values of Lee Weight}

\hspace*{0.6cm} We first recall the following classic lemmas, which plays an important role in determining the Lee weight of codewords of $C_m$.

\noindent\textbf{Lemma 4.1}\cite[(6) p.412]{MJ} If $\mathbf{y}=(y_1,y_2,\cdots,y_n)\in \mathbb{F}_2^n,$
then $2w_H(\mathbf{y})=n-\sum\limits_{i=1}^n(-1)^{y_i}.$\\

\noindent\textbf{Lemma 4.2}\cite[Lemma 9 p.143]{MJ} If $z \in \mathbb{F}_{2^m}^*,$ then $\sum\limits_{x\in \mathbb{F}_{2^m}}(-1)^{tr(z x)}=0.$

 We are now ready to discuss the Lee weight of the codewords in the above abelian codes.\\
\noindent\textbf{Theorem 4.3}\label{enum} For $a, b\in \mathcal{R}_m, m$ is odd, the Lee weight of the codewords of $C_m$ is given below.
\begin{enumerate}
{\item[(i)] If $a=0,b=0$, then $w_{L}(Ev(a,b))=0$;
\item[(ii)] If $b=0, a\neq0,$\\
1) $a\in M\backslash \{0\}$, then $w_{L}(Ev(a,b))=2^{2m}$;\\
2) $a\in \mathcal{R}_m^*$, then $w_{L}(Ev(a,b))=(2^{m}-1)2^m$;
\item[(iii)] If $a=0, b\neq0,$\\
1) $b\in M\backslash \{0\}$, then $w_{L}(Ev(a,b))=2^{2m}$;\\
2) $b\in \mathcal{R}_m^*$, then $w_{L}(Ev(a,b))=(2^{m}-1)2^m$;
\item[(iv)] If $a\neq0, b\neq0,$\\
1) $a\in M\backslash \{0\},b\in M\backslash \{0\}$, then $w_{L}(Ev(a,b))=2^{2m}, 2^{2m}-2^{\frac{3m+1}{2}}$ or $2^{2m}+2^{\frac{3m+1}{2}}$;\\
2) $a\in M\backslash \{0\}, b\in \mathcal{R}_m^*$, then $w_{L}(Ev(a,b))=(2^{m}-1)2^m$;\\
3) $a\in \mathcal{R}_m^*, b\in M\backslash \{0\}$, then $w_{L}(Ev(a,b))=(2^{m}-1)2^m$;\\
4) $a\in \mathcal{R}_m^*, b\in \mathcal{R}_m^*$, then $w_{L}(Ev(a,b))=2^m(2^m-2)$ or $2^{2m}$.
}
 \end{enumerate}
{\bf Proof.} (i) If $a=0,b=0$, then $Ev(a,b)=(\underbrace{0,0,\cdots,0}\limits_{|\mathcal{R}_m^*|})$. So $w_{L}(Ev(a,b))=0$.\\
(ii) If $b=0, a\neq0,$

 1) $a\in M\backslash \{0\}$, let $a=\beta u, \beta\in \mathbb{F}_{2^m}^*,~x=x_0+x_1u\in \mathcal{R}_m^*, x_0\in\mathbb{F}_{2^m}^*$. So we have $ax=\beta x_0u, Tr(ax)=tr(\beta x_0)u$. Taking Gray map yields $$\Phi(Ev(a,b))=(tr(\beta x_0),tr(\beta x_0))_{x_0,x_1}.$$
Using Lemma 4.1, we have $$2|\mathcal{R}_m^*|-2w_{L}(Ev(a,b))=2\sum_{x_0\in \mathbb{F}_{2^m}^*}~\sum_{x_1\in \mathbb{F}_{2^m}}(-1)^{ (\beta x_0)}=-2^{m+1}.$$
Then $w_{L}(Ev(a,b))=|\mathcal{R}_m^*|+2^m=2^{2m}.$

2) $a\in \mathcal{R}_m^*$, let $a=\alpha+\beta u\in \mathcal{R}_m^*,~x=x_0+x_1u\in \mathcal{R}_m^*$. So we have $ax=(\alpha+\beta u)(x_0+x_1u)=\alpha x_0+(\alpha x_1+\beta x_0)u, Tr(ax)=tr(\alpha x_0)+tr(\alpha x_1+\beta x_0)u$. Taking Gray map yields $$\Phi(Ev(a,b))=(tr(\alpha x_1+\beta x_0),tr(\alpha x_0)+tr(\alpha x_1+\beta x_0))_{x_0,x_1}.$$
From Lemma 4.1, we have\begin{eqnarray*}
                          2|\mathcal{R}_m^*|-2w_{L}(Ev(a,b)) &=& \sum_{x_0\in \mathbb{F}_{2^m}^*}~\sum_{x_1\in \mathbb{F}_{2^m}}(-1)^{ tr(\alpha x_1+\beta x_0)}+\\
                          &&\sum_{x_0\in \mathbb{F}_{2^m}^*}~\sum_{x_1\in \mathbb{F}_{2^m}}(-1)^{ tr(\alpha x_0)+tr(\alpha x_1+\beta x_0)} \\
                          &=& 0.
                        \end{eqnarray*}
Then $w_{L}(Ev(a,b))=|\mathcal{R}_m^*|=(2^{m}-1)2^m.$\\
(iii) We can use a similar approach as above to prove it, and omit here.\\
(iv) If $a\neq0, b\neq0,$

1) $a\in M\backslash \{0\},b\in M\backslash \{0\}$, let $a=\beta_1u, b=\beta_2u, \beta_1, \beta_2\in \mathbb{F}_{2^m}^*, x=x_0+x_1u\in R_m^*.$ So we have\begin{eqnarray*}
            ax+bx^3 &=& \beta_1u(x_0+x_1u)+\beta_2u(x_0+x_1u)^3 \\
             &=& \beta_1ux_0+\beta_2ux_0^3 \\
             &=& (\beta_1x_0+\beta_2x_0^3)u,
          \end{eqnarray*}
$$Tr(ax+bx^3)=tr(\beta_1x_0+\beta_2x_0^3)u.$$
Taking Gray map yields $$\Phi(Ev(a,b))=(tr(\beta_1x_0+\beta_2x_0^3),tr(\beta_1x_0+\beta_2x_0^3))_{x_0,x_1}.$$
 Combined with Lemma 4.1 and Lemma 4.2, we have $$2|\mathcal{R}_m^*|-2w_{L}(Ev(a,b))=2\sum_{x_0\in \mathbb{F}_{2^m}^*}~\sum_{x_1\in \mathbb{F}_{2^m}}(-1)^{ tr(\beta_1x_0+\beta_2x_0^3)}=2^{m+1}\sum_{x_0\in \mathbb{F}_{2^m}^*}(-1)^{ tr(\beta_1x_0+\beta_2x_0^3)}.$$ Let $A=\sum\limits_{x_0\in \mathbb{F}_{2^m}}(-1)^{ tr(\beta_1x_0+\beta_2x_0^3)}$. According to the proof of Theorem 10.1 in \cite[p.337]{SWG}, we can calculate the value of $A$. By squaring $A$, we have
\begin{eqnarray*}
  A^2 &=& \sum\limits_{x_0,\omega_0\in \mathbb{F}_{2^m}}(-1)^{ tr(\beta_1x_0+\beta_2x_0^3)+tr(\beta_1(x_0+\omega_0)+\beta_2(x_0+\omega_0)^3)} \\
   &=& \sum\limits_{x_0,\omega_0\in \mathbb{F}_{2^m}}(-1)^{ tr(\beta_1\omega_0+\beta_2\omega_0^3)+tr(\beta_2x_0\omega_0^2+\beta_2\omega_0x_0^2)}.
\end{eqnarray*}
Let $h(\omega_0)=tr(\beta_1\omega_0+\beta_2\omega_0^3)$, using the trace function identity $tr(\beta_2\omega_0 x_0^2)=tr(\sqrt{\beta_2\omega}_0x_0)$, we get
\begin{eqnarray*}
  A^2 &=& \sum\limits_{\omega_0\in \mathbb{F}_{2^m}}(-1)^{h(\omega_0)}\sum\limits_{x_0\in \mathbb{F}_{2^m}}(-1)^{tr(\beta_2x_0\omega_0^2+\beta_2\omega_0x_0^2)} \\
   &=& \sum\limits_{\omega_0\in \mathbb{F}_{2^m}}(-1)^{h(\omega_0)}\sum\limits_{x_0\in \mathbb{F}_{2^m}}(-1)^{tr((\beta_2\omega_0^2+\sqrt{\beta_2\omega_0})x_0)}.
\end{eqnarray*}

a) If $\beta_2\omega_0^2+\sqrt{\beta_2\omega_0}\neq0$, then $\sum\limits_{x_0\in \mathbb{F}_{2^m}}(-1)^{tr((\beta_2\omega_0^2+\sqrt{\beta_2\omega_0})x_0)}=0$.

b) If $\beta_2\omega_0^2+\sqrt{\beta_2\omega_0}=0$, we get $\omega_0(\beta_2\omega_0^3-1)=0$. Thus in the case $\omega_0=0,$ we have $h(\omega_0)=0$ and $(-1)^{h(\omega_0)}=1$. If $\beta_2\omega_0^3=1$, then $h(\omega_0)=tr(\beta_1\omega_0)+tr(1)=tr(\frac{\beta_1}{\sqrt[3]{\beta_2}})+1$ and $(-1)^{h(\omega_0)}=\pm1$. Since we know the value of $tr(\frac{\beta_1}{\sqrt[3]{\beta_2}})$ is $0$ or $1$. Thus
\begin{eqnarray*}
  A^2 &=& \sum\limits_{\omega_0=0,\frac{1}{\sqrt[3]{\beta_2}}}(-1)^{h(\omega_0)}\sum\limits_{x_0\in \mathbb{F}_{2^m}}(-1)^{tr((\beta_2\omega_0^2+\sqrt{\beta_2\omega_0})x_0)}+  \\
  && \sum\limits_{\omega_0\in \mathbb{F}_{2^m}\backslash\{0,\frac{1}{\sqrt[3]{\beta_2}}\}}(-1)^{h(\omega_0)}\sum\limits_{x_0\in \mathbb{F}_{2^m}}(-1)^{tr((\beta_2\omega_0^2+\sqrt{\beta_2\omega_0})x_0)}\\
   &=& 2^m\pm2^m+0 \\
   &=& 0 ~~~~{\rm or} ~~~~ 2^{m+1}.
\end{eqnarray*}It means that $A$ take three values $0, \pm2^{\frac{m+1}{2}}$, which implies $\sum_{x_0\in \mathbb{F}_{2^m}^*}(-1)^{ tr(\beta_1x_0+\beta_2x_0^3)}$ equals to $-1, \pm2^{\frac{m+1}{2}}-1$. Therefore we get $w_{L}(Ev(a,b))=2^{2m}, 2^{2m}\pm2^{\frac{3m+1}{2}}$.

2) $a\in M\backslash \{0\}, b\in \mathcal{R}_m^*$, let $a=\beta_1u, \beta_1\in \mathcal{R}_m^*, b=\alpha_2+\beta_2u\in \mathcal{R}_m^*,~x=x_0+x_1u\in \mathcal{R}_m^*$. So we have
\begin{eqnarray*}
  ax+bx^3 &=& \beta_1u(x_0+x_1u)+(\alpha_2+\beta_2u)(x_0+x_1u)^3 \\
   &=& \alpha_2 x_0^3+(\beta_1x_0+\alpha_2 x_0^2x_1+\beta_2x_0^3)u,\\
   Tr(ax+bx^3)&=& tr(\alpha_2 x_0^3)+tr(\beta_1x_0+\alpha_2 x_0^2x_1+\beta_2x_0^3)u.
\end{eqnarray*}
Taking Gray map yields $$\Phi(Ev(a,b))=(tr(\beta_1x_0+\alpha_2 x_0^2x_1+\beta_2x_0^3),tr(\alpha_2 x_0^3)+tr(\beta_1x_0+\alpha_2 x_0^2x_1+\beta_2x_0^3))_{x_0,x_1}.$$ In the light of Lemma 4.1, we have
\begin{eqnarray*}
  2|\mathcal{R}_m^*|-2w_{L}(Ev(a,b)) &=& \sum_{x_0\in \mathbb{F}_{2^m}^*}~\sum_{x_1\in \mathbb{F}_{2^m}}(-1)^{ tr(\beta_1x_0+\alpha_2 x_0^2x_1+\beta_2x_0^3)}+ \\
   && \sum_{x_0\in \mathbb{F}_{2^m}^*}~\sum_{x_1\in \mathbb{F}_{2^m}}(-1)^{tr(\alpha_2 x_0^3)+tr(\beta_1x_0+\alpha_2 x_0^2x_1+\beta_2x_0^3)}\\
   &=&0.
\end{eqnarray*}
Then $w_{L}(Ev(a,b))=|\mathcal{R}_m^*|=2^{m}(2^m-1)$.

3) Now we deal with the case $a\in \mathcal{R}_m^*$ and $b\in M\backslash \{0\}$. Deduce from computing $ax+bx^3$ that $ Tr(ax+bx^3)=tr(\alpha_1x_0)+tr(\alpha_1x_1+\beta_1x_0+\beta_2x_0^3)u$ with $a=\alpha_1+\beta_1u\in \mathcal{R}_m^*, b=\beta_2u, \beta_2\in \mathbb{F}_{2^m}^*,x=x_0+x_1u\in \mathcal{R}_m^*.$
Taking Gray map yields $$\Phi(Ev(a,b))=(tr(\alpha_1x_1+\beta_1x_0+\beta_2x_0^3),tr(\alpha_1x_0)+tr(\alpha_1x_1+\beta_1x_0+\beta_2x_0^3))_{x_0,x_1}.$$
 According to Lemma 4.1, we have \begin{eqnarray*}
  2|\mathcal{R}_m^*|-2w_{L}(Ev(a,b)) &=& \sum_{x_0\in \mathbb{F}_{2^m}^*}~\sum_{x_1\in \mathbb{F}_{2^m}}(-1)^{ tr(\alpha_1x_1+\beta_1x_0+\beta_2x_0^3)}+ \\
   && \sum_{x_0\in \mathbb{F}_{2^m}^*}~\sum_{x_1\in \mathbb{F}_{2^m}}(-1)^{tr(\alpha_1x_0)+tr(\alpha_1x_1+\beta_1x_0+\beta_2x_0^3)}\\
   &=&0.
\end{eqnarray*}
Then $w_{L}(Ev(a,b))=|\mathcal{R}_m^*|=2^{m}(2^m-1)$;

4) $a\in \mathcal{R}_m^*, b\in \mathcal{R}_m^*$, let $a=\alpha_1+\beta_1u\in \mathcal{R}_m^*,b=\alpha_2+\beta_2u\in \mathcal{R}_m^*,x=x_0+x_1u\in \mathcal{R}_m^*$. So we have
\begin{eqnarray*}
  ax+bx^3 &=&(\alpha_1+\beta_1u)(x_0+x_1u)+(\alpha_2+\beta_2u)(x_0+x_1u)^3 \\
   &=& (\alpha_1x_0+\alpha_2x_0^3)+(\alpha_1x_1+\beta_1x_0+\alpha_2x_0^2x_1+\beta_2x_0^3)u,  \\
 Tr(ax+bx^3) &=& tr(\alpha_1x_0+\alpha_2x_0^3)+tr(\alpha_1x_1+\beta_1x_0+\alpha_2x_0^2x_1+\beta_2x_0^3)u.
\end{eqnarray*}
Taking Gray map yields $\Phi(Ev(a,b))=(t r(\alpha_1x_1+\beta_1x_0+\alpha_2x_0^2x_1+\beta_2x_0^3),tr(\alpha_1x_0+\alpha_2x_0^3)+tr(\alpha_1x_1+\beta_1x_0+\alpha_2x_0^2x_1+\beta_2x_0^3))_{x_0,x_1}.$
Using Lemma 4.1, we have \begin{eqnarray*}
  2|\mathcal{R}_m^*|-2w_{L}(Ev(a,b)) &=& \sum_{x_0\in \mathbb{F}_{2^m}^*}~\sum_{x_1\in \mathbb{F}_{2^m}}(-1)^{ tr(\alpha_1x_1+\beta_1x_0+\alpha_2x_0^2x_1+\beta_2x_0^3)}+ \\
   && \sum_{x_0\in \mathbb{F}_{2^m}^*}~\sum_{x_1\in \mathbb{F}_{2^m}}(-1)^{tr(\alpha_1x_0+\alpha_2x_0^3)+tr(\alpha_1x_1+\beta_1x_0+\alpha_2x_0^2x_1+\beta_2x_0^3)}.
\end{eqnarray*}
Let \begin{eqnarray*}
      B &=& \sum\limits_{x_0\in \mathbb{F}_{2^m}^*}~\sum\limits_{x_1\in \mathbb{F}_{2^m}}(-1)^{ tr(\alpha_1x_1+\beta_1x_0+\alpha_2x_0^2x_1+\beta_2x_0^3)} \\
      &=& \sum\limits_{x_0\in \mathbb{F}_{2^m}^*}(-1)^{tr(\beta_1x_0+\beta_2x_0^3)}\sum\limits_{x_1\in \mathbb{F}_{2^m}}(-1)^{tr((\alpha_1+\alpha_2x_0^2)x_1)}.
    \end{eqnarray*}
If $\alpha_1+\alpha_2x_0^2=0,$ i.e., $x_0=\sqrt{\frac{\alpha_1}{\alpha_2}},$ thus $(-1)^{tr(\beta_1x_0+\beta_2x_0^3)}=\pm1$. If $\alpha_1+\alpha_2x_0^2\neq0,$ thus $\sum\limits_{x_1\in \mathbb{F}_{2^m}}(-1)^{tr((\alpha_1+\alpha_2x_0^2)x_1)}=0$. Therefore
\begin{eqnarray*}
  B&=& \sum\limits_{x_0=\sqrt{\frac{\alpha_1}{\alpha_2}}}(-1)^{tr(\beta_1x_0+\beta_2x_0^3)}~\sum\limits_{x_1\in \mathbb{F}_{2^m}}(-1)^{tr((\alpha_1+\alpha_2x_0^2)x_1)}+  \\
  &&\sum\limits_{x_0\in \mathbb{F}_{2^m}^*\backslash\sqrt{\frac{\alpha_1}{\alpha_2}}}(-1)^{tr(\beta_1x_0+\beta_2x_0^3)}~\sum\limits_{x_1\in \mathbb{F}_{2^m}}(-1)^{tr((\alpha_1+\alpha_2x_0^2)x_1)}\\
  &=& 2^m(\pm1)+0 \\
   &=& \pm2^m.
\end{eqnarray*}
Similarly, we can calculate the $\sum_{x_0\in \mathbb{F}_{2^m}^*}~\sum_{x_1\in \mathbb{F}_{2^m}}(-1)^{tr(\alpha_1x_0+\alpha_2x_0^3)+tr(\alpha_1x_1+\beta_1x_0+\alpha_2x_0^2x_1+\beta_2x_0^3)}=\pm2^m.$
Then $w_{L}(Ev(a,b))=|\mathcal{R}_m^*|=2^m(2^m-2)$, or $2^{2m}$.\\
Hence the theorem is proved.\qed

In the following, we give some concrete examples.\\
\noindent{\bf Example 4.4.} Let $m=3$. Then we obtain a binary code of parameters $[112,12,32]$. The weights are $\{32,48,56,64,96\}$.

\noindent{\bf Example 4.5.} Let $m=5$. Then we obtain a binary code of parameters $[1984,20,768]$. The weights are $\{768,960,992,1024,1280\}$.

%%%%%%%%%%%%%%%%%%%%%%%%%%%%%%%%%%%%%%%%%%%%%%%%%%%%%%%%%%%%%%%%%%%%%%%%%%%%%%%%%%%%%%%%%%%%%%%%%%%%%%%%%%%%%%%%%%%%%%%%%%%%%%%%%%%%%%%%%%%%%%%%%%%%%%%%%%%%%
\section{ The minimum distance of dual code}
\hspace*{0.6cm}We study the dual distance of $C_m$ in this section. However, we first investigate a property of trace function, that is nondegenerate.

\noindent{\bf Lemma 5.1}
 If for all $a, b \in \mathcal{R}_m,$ we have that $Tr(ax+bx^3)=0,$ then $x=0.$ \\
{\bf Proof.} Let $a=\alpha_1+\beta_1u, b=\alpha_2+\beta_2u$ and $x=x_0+x_1u$, where $x$ is a fixed element of $R$ and $\alpha_i,\beta_i, x_i \in \mathbb{F}_{2^m}, i=1,2.$ Then
\begin{eqnarray*}
  ax+bx^3 &=& (\alpha_1x_0+\alpha_2x_0^3)+(\alpha_1x_1+\beta_1x_0+\alpha_2x_0^2x_1+\beta_2x_0^3)u\\
   &=:& D_0+D_1u.
\end{eqnarray*}
 Thus $Tr(ax+bx^3)=0$ is equivalent to $tr(D_i)=0,i=0,1$. Using the nondegenerate character of $tr()$, we first have $tr(D_0)=0$, then we get $x_0=0$. Next we take $x_0=0$ into $tr(D_1)=0$, and we have $x_1=0$. Thus $x=0$. This completes the proof. \qed

Next, we give the dual Lee distance of the five-weight codes $C_m$ by use of Lemma 5.1.\\
\noindent\textbf{Theorem 5.2}
 For any positive integer $m,$ the dual Lee distance $d'$ of $C_m$ is $2.$\\
{\bf Proof.} We just need to show that $C_m^\perp$ does not contain codeword that has Lee weight $1$, but  contain codeword that has Lee weight $2$. Let $a=\alpha_1+\beta_1u, b=\alpha_2+\beta_2u,$ and $x=x_0+x_1u,y=y_0+y_1u$, $\alpha_i,\beta_i,x_i,y_i\in\mathbb{ F}_{2^m}, i=0,1$ and $x_0, y_0\neq 0.$

(1) If $C_m^\perp$ contains a codeword of Lee weight $1$, then there is a codeword of such that nonzero component of codeword has two types, that is \{$1$\}, \{$1+u$\}. This means that $\forall a, b\in \mathcal{R}_m$, $Tr(ax+bx^3)=0,$ or $(1+u)Tr(ax+bx^3)=0$ at some $x\in\mathcal{R}_m^*$, which implies
\begin{eqnarray*}
  Tr(ax+bx^3) &=& tr(\alpha_1x_0+\alpha_2x_0^3)+tr(\alpha_1x_1+\beta_1x_0+\alpha_2x_0^2x_1+\beta_2x_0^3)u \\
   &=& 0.
\end{eqnarray*}
Using Lemma 5.1, we conclude, by the nondegenerate character of $tr()$, that $x_0=0.$ Contradiction with $x_0\neq0$.

Hence, $C_m^\perp$ not contain a codeword of Lee weight $1$.

(2) If $C_m^\perp$ contains a codeword of Lee weight $2$, then there is a codeword of such that nonzero component of codeword has four types, that is $\{u\}, \{1,1\}, \{1,1+u\}$, or $\{1+u,1+u\}$. This implies that $\forall a, b\in \mathcal{R}_m$, $uTr(ax+bx^3)=0, Tr(ax+bx^3)+Tr(ay+by^3)=0, Tr(ax+bx^3)+(1+u)Tr(ay+by^3)=0$ or $(1+u)Tr(ax+bx^3)+(1+u)Tr(ay+by^3)=0$ at some $x,y\in\mathcal{R}_m^*$. In this case, $uTr(ax+bx^3)=0,$ by using similar method, we obtain contradiction.

We consider the case $Tr(ax+bx^3)+Tr(ay+by^3)=0$. By a simple calculation, we have a system of equations as follows:
\begin{equation*}
\begin{cases}
 x_0+y_0 &= 0;\\
 x_0^3+y_0^3 &= 0;\\
 x_1+y_1 &= 0;\\
  x_0^2x_1+y_0^2y_1 &= 0;
\end{cases}
\Longrightarrow
\begin{cases}
x_0&=y_0;\\
x_1&=y_1.
\end{cases}
\end{equation*}
Therefore, the number of $C_m^\perp$
contain a codeword of Lee weight $2$ with type $\{1,1\}$ is $(2^m-1)2^m$. Similarly, the number of $C_m^\perp$ contains codewords with type $\{1+u,1+u\}$ is $(2^m-1)2^{m}.$

From $Tr(ax+bx^3)+(1+u)Tr(ay+by^3)=0,$ we have a system of equations as follows:
\begin{equation*}
\begin{cases}
 x_0+y_0 &= 0;\\
 x_0^3+y_0^3 &= 0;\\
 x_1+y_1+y_0 &= 0;\\
  x_0^2x_1+y_0^2y_1+y_0^3 &= 0;
\end{cases}
\Longrightarrow
\begin{cases}
x_0&=y_0;\\
y_1&=x_0+x_1.
\end{cases}
\end{equation*}
Therefore, the number of $C_m^\perp$
contain a codeword of Lee weight $2$ with type $\{1,1+u\}$ is $(2^m-1)2^m$.

In the light of the above discussion, it is clearly to know that the number of $C_m^\perp$ contain a codeword of Lee weight $2$ is $3(2^m-1)2^m$, which implies the dual Lee distance $d'$ of $C_m$ is $2$. \qed

\section{ The Lee Weight distribution}
In this section, we calculate the Lee weight distribution of $C_m$, by using Pless power moments \cite{WC}.

According to Theorem 4.3, we have constructed a binary code of length $n=2^{m+1}(2^m-1)$, dimension $4m$, with five nonzero weights $\omega_1,~\omega_2,~\omega_3,~\omega_4$ and $\omega_5$~($\omega_1<\omega_2<\omega_3<\omega_4<\omega_5$) of values
\begin{eqnarray*}
  \omega_1 = 2^{2m}-2^{\frac{3m+1}{2}},~ \omega_2 =2^{m}(2^m-2), ~\omega_3= 2^{m}(2^m-1),~~\omega_4= 2^{2m},~~\omega_5= 2^{2m}+2^{\frac{3m+1}{2}},
\end{eqnarray*}
with respective frequencies $A_1,A_2,A_3,A_4,A_5$ denoted. Let $A_1^\perp,A_2^\perp, A_3^\perp, A_4^\perp$ denote frequencies of the Lee weight $1,2,3,4$ in $C_m^\perp$, respectively. In light of Theorem 5.2, we know $A_1^\perp=0,$ and $A_2^\perp=3(2^m-1)2^m.$

Now, we calculate the $A_3^\perp,A_4^\perp$. Let $a=\alpha_1+\beta_1u, b=\alpha_2+\beta_2u,$ $x=x_0+x_1u,y=y_0+y_1u, z=z_0+z_1u, t=t_0+t_1u$, $\alpha_i,\beta_i,x_i,y_i,z_i,t_i\in\mathbb{ F}_{2^m}, i=0,1$ and $x_0, y_0, z_0,t_0\neq 0.$

(1) If $C_m^\perp$ contains a codeword of Lee weight $3$, then there is a codeword such that nonzero component of codeword has six types, that is $\{1,u\}, \{1+u,u\},\{1,1,1\}, \{1,1,1+u\}, \{1,1+u,1+u\}$, or $\{1+u,1+u,1+u\}$. This means that $\forall a, b\in \mathcal{R}_m$, $Tr(ax+bx^3)+uTr(ay+by^3)=0, (1+u)Tr(ax+bx^3)+uTr(ay+by^3)=0, Tr(ax+bx^3)+Tr(ay+by^3)+Tr(az+bz^3)=0, Tr(ax+bx^3)+Tr(ay+by^3)+(1+u)Tr(az+bz^3)=0, Tr(ax+bx^3)+(1+u)Tr(ay+by^3)+(1+u)Tr(az+bz^3)=0$ or $(1+u)Tr(ax+bx^3)+(1+u)Tr(ay+by^3)+(1+u)Tr(az+bz^3)=0$ at some $x,y,z\in\mathcal{R}_m^*.$

In the case $Tr(ax+bx^3)+uTr(ay+by^3)=0,$ according to Lemma 5.1, we have $\alpha_1x_0+\alpha_2x_0^3=0$, then  $x_0=0$, contradiction with $x_0\neq0$. By using similar method, for $(1+u)Tr(ax+bx^3)+uTr(ay+by^3)=0$, we can obtain contradiction.

For case
$Tr(ax+bx^3)+Tr(ay+by^3)+Tr(az+bz^3) \\
~~~~~~~~~~~~~~~~~= tr(\alpha_1(x_0+y_0+z_0)+\alpha_2(x_0^3+y_0^3+z_0^3))
   +tr(\alpha_1(x_1+y_1+z_1)+\alpha_2(x_0^2x_1+\\~~~~~~~~~~~~~~~~~~~~~y_0^2y_1
   +z_0^2z_1)+\beta_1(x_0+y_0+z_0)+\beta_2(x_0^3+y_0^3
   +z_0^3))u\\
   ~~~~~~~~~~~~~~~~~=0.$

We can get $x_0+y_0+z_0=0$ and $x_0^3+y_0^3+z_0^3=0$. Using Newton's Identities, $x_0y_0z_0=0$, contradiction with $x_0, y_0,z_0\neq0$. For the balance of the three cases, we use a similar approach, drawing a contradictory.

From what has been discussed above, $C_m^\perp$ not contain a codeword of Lee weight $3$, i.e. $A_3^\perp=0$.

(2) If $C_m^\perp$ contain a codeword of Lee weight $4$, there is a codeword such that nonzero component of codeword has nine types, that is  $\{1,1,1,1\}, \{1+u,1+u,1+u,1+u\}, \{1,1+u,1+u,1+u\}, \{1,1,1+u,1+u\}, \{1,1,1,1+u\}, \{u,u\}, \{1,1,u\}, \{1,1+u,u\}$, or $\{1+u,1+u,u\}$.

1) For type $\{1,1,1,1\}$, we have $Tr(ax+bx^3)+Tr(ay+by^3)+Tr(az+az^3)+Tr(at+bt^3)=Tr(a(x+y+z+t)+b(x^3+y^3+z^3+t^3))=0$. Using Newton¡¯s Identities, $(X-x)(X-y)(X-z)(X-t)=X^4+\sigma_1X^3+\sigma_2X^2+\sigma_3X+\sigma_4=0.$ Thus we obtain $\sigma_1=x+y+z+t=0, S_3=x^3+y^3+z^3+t^3=0, \sigma_3=0,$ i.e., $X^4+\sigma_2X^2+\sigma_4=0.$ Taking $\omega=X^2,$ then $\omega^2+\sigma_2\omega+\sigma_4=0$ have two solutions if and only $X^4+\sigma_2X^2+\sigma_4=0$ has two solutions, which contradiction with $X^4+\sigma_1X^3+\sigma_2X^2+\sigma_3X+\sigma_4=0$ has four different solutions. Therefore, $C_m^\perp$ not contain a codeword of Lee weight $4$ with type $\{1,1,1,1\}$.
Similar we can prove that several cases with $\{1+u,1+u,1+u,1+u\}, \{1,1+u,1+u,1+u\}, \{1,1,1+u,1+u\}, \{1,1,1,1+u\}$ are impossible.

2) For type $\{u,u\}$, we have $uTr(ax+bx^3)+uTr(ay+by^3)=utr(\alpha_1(x_0+y_0)+\alpha_2(x_0^3+y_0^3))=0,$ then
\begin{equation*}
\begin{cases}
 x_0+y_0 &= 0;\\
 x_0^3+y_0^3 &= 0;
\end{cases}
\Longrightarrow x_0=y_0.
\end{equation*}
Therefore, the number of $C_m^\perp$ contain a codeword of Lee weight $4$ with type $\{u,u\}$ has $(2^m-1)2^{2m}.$

3) For type $\{1,1,u\}$, we have $Tr(ax+bx^3)+Tr(ay+by^3)+uTr(az+az^3)=tr(\alpha_1(x_0+y_0)+\alpha_2(x_0^3+y_0^3))+utr(\alpha_1(x_1+y_1+z_0)+\alpha_2(x_0^2x_1+y_0^2y_1+z_0^3)+\beta_1(x_0+y_0)+\beta_2(x_0^3+y_0^3))=0,$ then\begin{equation*}
\begin{cases}
 x_0+y_0 &= 0;\\
 x_0^3+y_0^3 &= 0;\\
 x_1+y_1+z_0 &= 0;\\
 x_0^2x_1+y_0^2y_1+z_0^3 &= 0;
\end{cases}
\Longrightarrow \begin{cases}
 x_0 &=y_0=z_0;\\
 y_1 &=x_0+x_1;
\end{cases}
\end{equation*}
Therefore, the number of $C_m^\perp$ contain a codeword of Lee weight $4$ with value $\{1,1,u\}$ has $(2^m-1)2^{2m}.$
Similar we can prove that case with $\{1+u,1+u,u\}$ has $(2^m-1)2^{2m}.$

4) For type $\{1,1+u,u\}$, we have $Tr(ax+bx^3)+(1+u)Tr(ay+by^3)+uTr(az+az^3)=tr(\alpha_1(x_0+y_0)+\alpha_2(x_0^3+y_0^3))+utr(\alpha_1(x_1+y_1+y_0+z_0)+\alpha_2(x_0^2x_1+y_0^2y_1+y_0^3+z_0^3)+\beta_1(x_0+y_0)+\beta_2(x_0^3+y_0^3))=0,$ then\begin{equation*}
\begin{cases}
 x_0+y_0 &= 0;\\
 x_0^3+y_0^3 &= 0;\\
 x_1+y_1+y_0+z_0 &= 0;\\
 x_0^2x_1+y_0^2y_1+y_0^3+z_0^3 &= 0;
\end{cases}
\Longrightarrow \begin{cases}
 x_0 &=y_0=z_0;\\
 x_1 &=y_1;
\end{cases}
\end{equation*}
Therefore, the number of $C_m^\perp$ contain a codeword of Lee weight $4$ with value $\{1,1+u,u\}$ has $(2^m-1)2^{2m}.$

To sum up, we have $A_4^\perp=4(2^m-1)2^{2m}=(2^m-1)2^{2m+2}.$

In order to calculate the frequencies $A_1,A_2,A_3,A_4,A_5$, using Pless power moments, we have
\begin{equation*}
\begin{cases}
 A_1+A_2+A_3+A_4+A_5 &= 2^{4m}-1;\\
\omega_1A_1+\omega_2A_2+\omega_3A_3+\omega_4A_4+\omega_5A_5 &= 2^{4m-1}n ;\\
\omega_1^2A_1+\omega_2^2A_2+\omega_3^2A_3+\omega_4^2A_4+\omega_5^2A_5 &= 2^{4m-2}(n(n+1)+2A_2^\perp);\\
\omega_1^3A_1+\omega_2^3A_2+\omega_3^3A_3+\omega_4^3A_4+\omega_5^3A_5 &= 2^{4m-3}(n^2(n+3)+6nA_2^\perp) \\
  \omega_1^4A_1+\omega_2^4A_2+\omega_3^4A_3+\omega_4^4A_4+\omega_5^4A_5 &= 2^{4m-4}(n(n+1)(n^2+5n-2)+4(3n^2+3n-4)A_2^\perp+24A_4^\perp)
\end{cases}
\end{equation*}
\textbf{Remark:} \textbf{\color{red}When we calculate the above equations, we find the values of $A_1, A_2, A_3, A_4, A_5$ are not integers. It indicates that there must be some mistakes in them. If you are interested in this problem, can you find it and tell me ? We are pleased to cooperate with you. }

\section{Application to secret sharing schemes}
In this section, we first introduce the support structure.
Let $q$ be a prime power, and $n$ an integer. Let $\mathbb{F}_q$ denote the finite field of order $q.$
The \textbf{support} $s(x)$ of a vector $x$ in $\mathbb{F}_q^n$ is defined as the set of indices where it is nonzero. We say that a vector $x$ covers a vector $y$ if $s(x)$ contains $s(y).$
A \textbf{minimal codeword} of a linear code $C$ is a nonzero codeword that does not cover any other nonzero codeword. In general determining the minimal codewords
of a given linear code is a difficult task. However, there is a numerical condition, derived in \cite{AB}, bearing on the weights of the code, that is easy to check.

\noindent{\bf Lemma 7.1 }(Ashikmin-Barg) Denote by $w_0$ and $w_{\infty}$ the minimum and maximum nonzero weights, respectively. If
$$\frac{w_0}{w_{\infty}}>\frac{q-1}{q},$$ then every nonzero codeword of $C$ is minimal.

We can infer from there the support structure for the codes of this paper.

\noindent{\bf Proposition 7.2}
 All the nonzero codewords of $\Phi(C_m),$ for $m> 2$ and $m$ is odd, are minimal.\\
{\bf Proof.}
 Based on the introduction of Lemma 6.1, then $w_0=\omega_1,$ $w_{\infty}=\omega_5$ and $q=2.$ Next we need to prove the inequality $\frac{w_1}{w_{5}}>\frac{1}{2}$ is true for $m> 2$. Thus, we obtain
 \begin{eqnarray*}
  2\omega_1-\omega_5&=& 2(2^{2m}-2^{\frac{3m+1}{2}})-(2^{2m}+2^{\frac{3m+1}{2}}) \\
    &=& 2^{2m}(1-3\cdot2^{1-m})>0.
 \end{eqnarray*}
Hence the proposition is proved.\qed

A secret sharing scheme (SSS) is a protocol involving a dealer and $U$ users.
\textbf{Massey's scheme} is a construction of such a scheme where a code $C$ of length $n$ over $\mathbb{F}_p$ gives rise to a SSS with $U=n-1.$
See \cite{YJ} for a detailed explanation
of the mechanism of that scheme. Now, the coalition structure is related to the support
structure of $C.$ In the special case when all nonzero codewords are minimal, it was shown in \cite{YC} that there is the following alternative, depending on $d'$:
\begin{itemize}
 \item If $d'\ge 3,$ then the SSS is \emph{``democratic''}: every user belongs to the same number of coalitions.
 \item If $d'=2,$  then there are users  who belong to every coalition: the \emph{``dictators''.}
\end{itemize}
Depending on the application, one or the other situation might be more suitable.
By Proposition 7.2, and Theorem 5.2, we see that a SSS built on $\Phi(C_m)$ is dictatorial.

%On the other hand, it is easy to check that the results are correct by listing the generator matrices of $Q_{1},S_{1}$ and $ \overline{Q}_{1}.$

\section{Conclusion}
\hspace*{0.6cm}In this paper, we have studied a family of trace codes over $\mathbb{F}_2+u\mathbb{F}_2.$ These codes we construct are provably abelian, but not visibly cyclic. Using a character sum approach, Pless power moments and Newton¡¯s Identities, we have been able to determine their Lee weight distribution of $C_m$, and we have obtained a family of abelian binary five-weight codes by the Gray map.

%A metric study, including Gray map, and distances of the $p$-ary image of codes of reasonable length is worth considering for further study.
%Therefore, the study of quadratic residue codes over $\F_p+v\F_p+v^{2}\F_p$ has a certain significance.

%\end{CJK}

%\end{document}

\end{document}